\documentstyle[12pt,psfig]{article}
\newcommand{\be}{\begin{equation}}
\newcommand{\ee}{\end{equation}}
\newcommand{\ba}{\begin{eqnarray}}
\newcommand{\ea}{\end{eqnarray}}

\begin{document}

{\pagestyle{empty}
                                             \begin{flushright}
                                             SPbU-IP-96-1 \\
                                             hep-ph/9612xxx
                                             \end{flushright}

\vskip 3cm
\centerline{ \large\bf
              Cross-Symmetric
                Expansion of $\pi \pi$ Amplitude
                Near Threshold
           }
\vskip 1cm
\centerline{ \bf
              A.A. Bolokhov, T.A. Bolokhov and I.S. Manida }
\centerline{ \it
 St. Petersburg State University, Sankt--Petersburg, Russia}
\centerline{ \bf
                         and }
\centerline{ \bf
                 M.V. Polyakov and S.G. Sherman }
\centerline{ \it
St. Petersburg Nuclear Physics Institute,
Sankt--Petersburg, Russia }
\vskip 2cm

\begin{abstract}

                The near--threshold expansion of the
        $ \pi \pi $
                amplitude is developed
                using the crossing--covariant independent
                variables.
                The independent threshold parameters
                entering the real part of the amplitude
                in an explicitly Lorentz--invariant way
                are free from restrictions of isotopic
                and crossing symmetries.
                Parameters of the expansion
                of the imaginary part
                are recovered by the perturbative
                unitarity relations.

\end{abstract}

\vskip 6cm
\centerline{
                     Sankt--Petersburg 1996 }

\newpage
}

{\large
\centerline{
              Cross-Symmetric
                Expansion of $\pi \pi$ Amplitude
                Near Threshold
           }
}
\vskip0.3cm
\centerline{ \bf
                A.A.~Bolokhov, T.A.~Bolokhov, I.S.~Manida }
\centerline{\sl Sankt-Petersburg State University,
                Sankt-Petersburg, 198904, Russia }

\centerline{ \bf
                 M.V.~Polyakov, S.G.~Sherman }
\centerline{ \sl
St. Petersburg Institute for Nuclear Physics,
Sankt--Petersburg, 188350, Russia }
\vskip 0.5cm

\centerline{\bf Abstract }
\vskip0.2cm

\noindent
                The near--threshold expansion of the
        $ \pi \pi $
                amplitude is developed
                using the crossing--covariant independent
                variables.
                The independent threshold parameters
                entering the real part of the amplitude
                in an explicitly Lorentz--invariant way
                are free from restrictions of isotopic
                and crossing symmetries.
                Parameters of the expansion
                of the imaginary part
                are recovered by the perturbative
                unitarity relations.

\vskip1.cm
\section{ Introduction }

                The growth of interest to the pion interactions
                is currently motivated by the achievements
                of the Chiral
                Perturbation Theory (ChPT)
                which was formulated
                by Weinberg, Gasser and Leutwyler
\cite{SW79,GL8485}
                as an effective low energy limit of QCD.
                Already first successful
                one--loop calculations of the near--threshold
        $ \pi \pi $
                amplitude
                by M.~Volkov and Pervushin
\cite{MVolkovP7475}
                had shown the predictive power
                of Chiral Symmetry in the framework
                of approach
                of nonlinear effective lagrangians
                (see
\cite{MVolkovP78}).
                The importance of
                the threshold characteristics of the
        $ \pi \pi $
                scattering
                explains also the need in more precise
                experimental measurements
                and the model--independent analysis of
                these characteristics.


                During all the times being,
                the central role of the
        $ \pi \pi $
                scattering
                in the theory of strong interactions
                is steming from its remarkable properties
                which might be briefly collected
                into the following list:


                1.  Pions are the lightest hadronic states;
                unitarity provides that only the elastic
        $\pi\pi$
                cut is important at low energies.

                2. Pions are spinless particles: the amplitude
                assumes the simplest form.

                3. The isospin symmetry restricts
                the amplitudes
                of various physical processes to only 3
                independent combinations since
                the two--pion state might have only
                three values of the fixed
                isospin
        $ I = 2, \; 1, \; 0$.

                4. This reaction provides an example of
                the perfect crossing symmetry.
                The three isospin amplitudes are still not
                independent in the functional sense.
                Bose--statistics of pions considered as
                identical particles
                and crossing relations for
                amplitudes of cross channels allow to express
                any two amplitudes in terms of the rest one.

                5. The interpretation of the pion as the
                Goldstone boson of the spontaneously broken
                Chiral symmetry of QCD
                provided
                a new
                argumentation for the importance
                of the pion--pion
                interactions in the hadron physics and
                nuclear phenomena at low energies.


                The few listed properties
                explain why this reaction for a long time
                serves as a test field for various methods of
                the quantum field theory
                such as dispersion relations,
                the
        $S$--matrix
                approach, etc.
                The advantages and achievements of various
                theoretical approaches
                are summarized in the review books
\cite{MMS,BelkovMP85}
                --- we advise reader to look there
                for more details if necessary;
                the modern point of view on
                pion--pion and pion--hadron
                interactions
                might be found in the book
\cite{DGH}
                and details of ChPT approach
                --- in the review papers
\cite{UGM93,Pich95};
                a summary of the most interesting
                theoretical predictions as well as
                of the forthcoming experimental tests
                might be found in the talk
\cite{Pocanic94}.

                The recent results
\cite{KnechtMSF95}
                of the so called
                Generalized ChPT approach
\cite{JSternSF93}
                and the progress
                in the two--loop ChPT calculations
\cite{BijenesCEGS96,KnechtMSF95}
                are claiming for more precise
                experimental information on the
        $\pi\pi$
                interaction at low energies at the
        $ O ( k^{6} ) $
                order
                which obviously must be presented
                in the model--independent form.

                Since it is
                not possible
                experimentally
                to create the pionic target
                or the colliding pion beams
                there are only indirect ways
                for obtaining experimental data
                on the
        $ \pi \pi $
                scattering.
                The reactions
        $  \pi N \to \pi \pi N $
                and
        $  K \to \pi \pi e {\nu} $
                are considered as the most important sources
                of an (indirect) information.
                At the same time
                more extended variety of processes
                like
        $  \gamma N \to \pi \pi N $
                must rely upon the same properties
                of OPE mechanism
                and/or the final--state interaction of pions
                which are common to the description of
        $  \pi N \to \pi \pi N $
                and
        $  K \to \pi \pi e {\nu} $
                reactions.
                Therefore an unambiguous parametrization
                of the
        $  \pi \pi  $
                amplitude
        ($  4 \pi  $
                vertex)
                is required for the low energy region.

                In principle,
                the parameters of phase shifts
                or the scattering lengths and the slopes
                of partial waves
                are the most acceptable physical parameters.
                However,
                the partial wave decomposition
                can not respect the crossing properties
                in a simple way.
                The technique of
                Roy equations
\cite{Roy71}
                based on
                dispersion relations
                can restrict the
                parameters and make the amplitude
                (which is defined in
                terms of partial waves)
                consistent with
                conditions of Bose--statistics and crossing.
                The approach relies upon high energy
        $  \pi \pi  $
                data.
                Since the latter are neither infinitely precise
                nor free from contradictions
                only the bands for the scattering lengths
                had been provided
                in the most recent application of this approach
                in the paper
\cite{OOPTM96}.

                The need to provide the analysis of data on
                processes like
        $  \pi N \to \pi \pi N $,
        $  \gamma N \to \pi \pi N $
                where the amplitude of
        $  \pi \pi  $
                scattering must be considered
                in the presence of contributions
                of a variety of concurrent processes
                puts further restrictions on the desired
                parametrization of the
        $  4 \pi  $
                vertex.
                Since the analyzed amplitude comes with
                a large number of free parameters the fitting
                of experimental data becomes possible
                only under the condition
                that the phase--space integration
                might be factored out from parameters.
                This is possible only if free (and formal)
                parameters enter the amplitude polynomially
                and the relations for formal parameters
                do not contain kinematics;
                otherwise the huge number of the
                9--dimensional
                integration runs
                would be inevitable for all experimental points
                during the every step of fitting iterations.

                The condition rules out the phase shift
                parameters as well as
                the next--to--leading order ansatz of ChPT for
                the
        $  4 \pi  $
                vertex.

                The main goal of the present paper
                is to elaborate the near--threshold
                expansion of the
        $  \pi \pi $
                amplitude which along with
                the isospin invariance
                satisfies
                the exact combined Bose and crossing symmetries
                and the approximate
                (perturbative) unitarity.

                Our approach is based on the crossing covariant
                variables
                which were introduced in the paper
\cite{BolokhovVS88pi}
                and proved to be an efficient tool
                for the off--mass--shell parametrization of
                the considered amplitude,
                in particular, when being applied to the
        $  \pi N \to \pi \pi N $
                reaction
\cite{BVS}.
                In the present paper we use combinations
                of parameters
                describing the deviation from the threshold
                in the crossing--invariant way.
                In their terms the real part of the amplitude
                is uniformly expanded
                in the vicinities of the thresholds
                of all physical channels simultaneously.
                The region of validity of the expansion
                is in general restricted by the
                Mandelstam domain of analiticity
\cite{SMand60}
\begin{equation}
\label{OMD}
             | s t u | < 288 \mu^6
          \; .
\end{equation}

                The paper is organized as follows.
                The content of the sect. 2 reminds
                definitions and basic properties
                of the considered amplitude.
                In sect. 3 the analysis of crossing properties
                of the amplitude is provided on the base of
                crossing covariant variables
                of the paper
\cite{BolokhovVS88pi}.
                The results are used
                for the final construction
                of the low energy phenomenological amplitude
                in sect. 4.
                The summary, the concluding remarks
                and the  discussion on the field of
                possible applications are given in Conclusions.

      \section{  General Properties of
        $\pi\pi$
                Amplitude}

                We shall consider the amplitude
        $ M_{adbc}$
                of the auxiliary reaction
\begin{equation}
          \pi_a ( k_1 ) + \pi_d ( k_4 ) \rightarrow
                \pi_b ( k_2 ) + \pi_c ( k_3 )
                \; ,
\label{react}
\end{equation}
%
%
%
%
                and define it by
\begin{equation}
          \langle   \pi_b ( k_2 ) , \pi_c ( k_3 )
                \; | \;  S - 1  \; | \;
          \pi_a ( k_1 ) , \pi_d ( k_4 ) \rangle
                =  i ( 2 \pi )^4
                \delta^4 ( k_1 + k_4 - k_2 - k_3 )
                          M_{adbc}
                \; .
\label{Amp}
\end{equation}
                The amplitude of any physical
                process is obtained in terms of
        $ M_{adbc}$
                using the definitions of (charged) pion
                fields:
\begin{equation}
          \pi^{\pm}  = \frac{1}{\sqrt{2}}
              ( \pi_1 \pm i \pi_2 )
                                         \; , \;
                \pi^0  =  \pi_3  \; .
\label{pions}
\end{equation}

                The essence of the isotopic symmetry might be
                exploited in two ways. First, it allows
                to write
        $ M_{adbc}$
                in terms of the isospin projections (the direct
        $s$
                channel
                is selected here):
\begin{equation}
         M_{adbc} =
         P^{I=2}_{ad,bc} T^2
          + P^{I=1}_{ad,bc} T^1
          + P^{I=0}_{ad,bc} T^0
                 \; ,
\label{MProj}
\end{equation}
\begin{eqnarray}
\label{Proj}
         P^{I=2}_{ad,bc} =
                - \frac{1}{3} \delta_{ad} \delta_{bc}
                + \frac{1}{2} ( \delta_{bd} \delta_{ac}
                              + \delta_{cd} \delta_{ab} )
                 \; ; \;\;
         P^{I=1}_{ad,bc} =
                - \frac{1}{2} ( \delta_{bd} \delta_{ac}
                              - \delta_{cd} \delta_{ab} )
                 \; ; \;\;
         P^{I=0}_{ad,bc} =
                  \frac{1}{3}  \delta_{ad} \delta_{bc}
                 \; .
\end{eqnarray}

                Second, one can define
                the isoscalar amplitudes
        $A, B, C$:
\begin{equation}
         M_{adbc} =
                        A     \delta_{ad} \delta_{bc}
                +       B     \delta_{bd} \delta_{ac}
                +       C     \delta_{cd} \delta_{ab}
                 \; ,
\label{MABC}
\end{equation}
                related to the fixed--isospin amplitudes
(\ref{MProj})
                by
\begin{eqnarray}
         T^0  = 3 A  +  B  +  C
                 \; ; \; \;
         T^1  =      -  B  +  C
                 \; ; \; \;
         T^2  =         B  +  C
                 \; .
\label{FxdI}
\end{eqnarray}

                The kinematical variables are the usual
                Mandelstam ones
\begin{eqnarray}
\label{stu}
         s  = ( k_1 + k_4 )^2 = ( k_2 + k_3 )^2
                 \; ; \;\;
         t  = ( k_1 - k_2 )^2 = ( k_4 - k_3 )^2
                 \; ; \;\;
         u  = ( k_1 - k_3 )^2 = ( k_4 - k_2 )^2
                 \; ,
\end{eqnarray}
                which are restricted on the mass shell of the
        $ 4\pi $
                vertex by
\begin{equation}
                s + t + u =
                \sum_i \mu^2_i
                =  4 \mu^2
                 \; .
\label{hmass}
\end{equation}

                To study the crossing properties it is suitable
                to consider
        $A, B, C$
                as functions of all set
        $s, t, u$
        ($ A = A ( s, t, u ) $, etc.).
                Then the crossing properties of
                the amplitude combined
                with Bose--statistics of pions provide the
                conditions
\begin{eqnarray}
                A ( s, t, u ) = A ( s, u, t )
                 \; ; \;
                B ( s, t, u ) = B ( t, s, u )
                 \; ; \;
                C ( s, t, u ) = C ( u, t, s )
                 \; ,
\label{crc}
\end{eqnarray}
                and the relations
\begin{equation}
\nonumber
                B ( s, t, u ) = A ( u, s, t ) = C ( t, u, s )
                 \; .
\label{crr}
\end{equation}

                These very properties look much more involved
                when formulated in terms of the decomposition
(\ref{MProj}).
                The latter is suitable for defining physical
                (observable) characteristics.
                It makes the strong preference of the
       $s$
                channel,
                so the variables usually chosen as
                independent are
       $ s $
                and
       $ t - u $
                or the momentum
                in the Center of Mass Frame (CMF)
       $ | {\bf k } | $
                and the scattering angle
       $ {\Theta } $:
\begin{equation}
               s = 4 ( {\bf k}^2 + \mu^2 )
                                         \; , \;
               t - u = 4 {\bf k}^2 \cos \Theta
                                         \; .
\label{CMF}
\end{equation}

                Because of the residual symmetry in the
                decomposition
(\ref{MProj})
                the amplitudes
        $ T^0 $
                and
        $ T^2 $
                must be symmetric with respect to the
        $ ( t \leftrightarrow u ) $
                permutation
                and, hence, have only even powers of the
        $ ( t - u ) $
                variable, while
        $ T^1 $
                has only odd ones.

                At the energies below the
        $ 4 \pi $
                threshold only the elastic unitarity conditions
                are of importance.
                They are obtained by
                inserting the 2--pion intermediate state into
                the unitarity condition for the
        $ S $--matrix
        $ S = 1 + i T $:
\begin{equation}
                \frac{1}{i} ( T - T^{\dag} ) = T T^{\dag}
                \; .
\label{Suni}
\end{equation}

                Because of the isospin conservation
                the resulting equations are diagonal in
        $ T^I $.
                They assume simple algebraic form for
                the partial
                waves
        $ T^I_{L} $
                defined via
\begin{equation}
         T^I  = ( 32 \pi ) \sum_L ( 2 L + 1 )
                            P_L ( \cos \Theta )
                            T^I_L ( | {\bf k} | )
                \; ,
\label{PW}
\end{equation}
                or
\begin{equation}
         T^I_L  = \frac{1}{( 32 \pi )}
                  \frac{1}{ 2 }
          \int^1_{-1} d \cos ( \Theta )
                            P_L ( \cos \Theta )
                            T^I
                \; .
\label{rPW}
\end{equation}
                The elastic unitarity requires
\begin{equation}
         {\rm Im } ~T^I_L  = \sigma ( s )
                    [ ({\rm Re } ~T^I_L)^2
                    + ({\rm Im } ~T^I_L)^2 ]
                \; .
\label{PWu}
\end{equation}
                where
\begin{equation}
          \sigma ( s )
                    = \sqrt{ \frac{ s - 4 \mu^2 }
                                  {      s      }
                            }
                    = \sqrt{ \frac{ {\bf k}^2 }
                                  { {\bf k}^2 + \mu^2 }
                            }
\label{sig}
\end{equation}
                is the kinematical factor
                originating from the integration over
                the phase space of
                the intermediate 2--pion state.
                The solution of eqs.
(\ref{PWu})
                is provided usually
                in terms of the phase shifts
         $ \delta^I_L (s) $:
\begin{equation}
         T^I_L  = \frac{ e^{i \delta^I_L } \sin \delta^I_L }
                       { \sigma }
                = \frac{ e^{2i \delta^I_L } - 1 }
                       { 2 i \sigma }
                \; .
\label{PWsol}
\end{equation}

                Another suitable solution of relations
(\ref{PWu})
                is being written in terms of the
                {\it reaction amplitudes}
         $ \phi^I_L (s) $:
\begin{equation}
         T^I_L  = \frac{ \phi^I_L }
                       { 1 - i \sigma  \phi^I_L }
                = \frac{ \varphi^I_L }
                       { 1 - i {\bf k}^{2L} \sigma \varphi^I_L }
                               {\bf k}^{2L}
                \; ,
\label{RAmp}
\end{equation}
                where the threshold behavior of
         $ \phi^I_L (s) =
         {\bf k}^{2L} \varphi^I_L (s) $,
         $  \varphi^I_L (s_{0}) $
                being finite,
                follows from the threshold properties of
        $ {\bf k}^{2L + 1} \cot  \delta^I_L (s) $.

                The examination of the elastic unitarity
                conditions
(\ref{PWu})
                and their solutions
(\ref{RAmp})
                at the small momentum
                shows that

                1) the imaginary part
                of the amplitude vanishes at the threshold;

                2) while the real part admits a Taylor series
                expansion in invariant variables,
                the imaginary part --- does not:
                there is the nonanalytic multiplier
        $  \sigma ( s ) $
                in the eq.
(\ref{PWu});

                3) the same multiplier prevents the
                simultaneous expansion in the momentum
        $ {\bf k}^2 $
                and the pion mass
        $ \mu^2 $
                of both real and imaginary parts.
                This
                in short terms explains why the so called
                nonanalytic terms must be present in the
                amplitude derived in the
                Chiral Perturbation Theory
                beyond
                the tree approximation;

                4) it is easy to solve the eq.
(\ref{PWu})
                for
        $ {\rm Im} ~T^I_L $:
\begin{equation}
       {\rm Im} ~T^I_L =
       \frac { 1 - \sqrt{ 1 -
                          4 (\sigma {\rm Re} ~T^I_L)^2
                         }
              }
             { 2 \sigma }
                \; .
\label{Imsol}
\end{equation}

                Thus, near the threshold it suffices to know
                the coefficients of the momentum expansion
                of the real part.
                The expansion for the
        $ L $--th
                partial wave must
                start from the order
        $ {\bf k}^{2L} $ ---
                every cosine appears in an invariant variable
                being multiplied by
        $ {\bf k}^{2} $
                (see eqs.
(\ref{CMF})).

                Therefore, the scattering lengths
        $ a^I_L \equiv a^I_{L,0} $,
                the slopes
        $ a^I_{L,1} $
                as well as the slopes
        $ a^I_{L,m} $
                of the higher order
        $ m $
                are used to be defined as coefficients of
                the following expansion
\begin{equation}
       {\rm Re} ~T^I_L = {\bf k}^{2L}
             (  a^I_L +
                a^I_{L,1} {\bf k}^{2} +
                a^I_{L,2} {\bf k}^{4} +
                \dots )
                \equiv    {\bf k}^{2L} R^I_L
                \; .
\label{ReTl}
\end{equation}

                The above quantities
        $  a^{I}_{L,m} $
                appear as the threshold
                (or the near--threshold)
                characteristics of the
        $ \pi \pi $
                interaction.
                This provides common expectations
                that the most conclusive predictions
                of the Chiral Perturbation Theory
                should be just about
                these very characteristics.

                The solution
(\ref{Imsol})
                provides the following pattern
                for the expansion of the imaginary part

\begin{equation}
       {\rm Im} ~T^I_L = \sigma ( {\rm Re} ~T^I_L )^{2}
             [ 1 +
                (\sigma  {\rm Re} ~T^I_L )^{2} +
               2 (\sigma  {\rm Re} ~T^I_L )^{4} +
               5 (\sigma  {\rm Re} ~T^I_L )^{6} +
                \dots ]
                \; .
\label{ImTlRe}
\end{equation}

                Thus, the momentum expansion of
                the imaginary part goes like
\begin{equation}
       {\rm Im} ~T^I_L = \sigma {\bf k}^{4L}
             ( \beta^I_L +
                \beta^I_{L,1} {\bf k}^{2} +
                \beta^I_{L,2} {\bf k}^{4} +
                \dots )
                \; ,
\label{ImTl}
\end{equation}
                and its coefficients
        $ \beta^I_{L,n} $
                are completely determined by the coefficients
        $ a^I_{L,n} $
                of the real part by virtue of eq.
(\ref{Imsol}).

                The formulae collected in the present section
                will be used below for the analysis
                of the crossing properties
                of the considered amplitude
                and the elaboration of
                the phenomenological ansatz
                suitable for the threshold region.
      \section{ Crossing Covariant Expansion }


                For the purpose of the analysis of the
        $ \pi \pi $--amplitude
                properties induced by Bose--statistics
                and crossing relations
                it was found convenient to introduce the
        {\it independent} variables
\cite{BolokhovVS88pi}
                which respect the
                covariance under permutations of pions
                --- let us call them the
        {\bf crossing--covariant} variables.
                (Neither the set
        $\{  s, t \} $
                nor
        $ \{ s, t - u \}$
                can meet the covariance requirements.)
                In terms of the cubic roots of unity
        $  \epsilon, \bar{\epsilon} $:
\begin{equation}
         \epsilon \equiv \exp{ ( 2 \pi i / 3 )} \; ,
         \bar{\epsilon} \equiv \exp{ ( - 2 \pi i / 3 )} =
         \epsilon^{2} = \epsilon^{*} \;
\end{equation}
                the suitable variables are defined by
\begin{eqnarray}
\label{teate}
              \theta  =
                       ( - k_{1}
                        + {\epsilon} k_{2}
                        + \bar{\epsilon} k_{3}
                        ) \cdot k_{4}
        \; ; \; \;
              \bar{\theta }  =
                       ( - k_{1}
                        + \bar{\epsilon} k_{2}
                        + {\epsilon} k_{3}
                        ) \cdot k_{4}
        \; .
\end{eqnarray}
                The inverse relations
                for all scalar products of
                particle momenta read
\begin{eqnarray}
\nonumber
         k_{1} \!\cdot\! k_{4} = k_{2} \!\cdot\! k_{3}
               & = &
                       - \frac{1}{3}
                      [
                        \mu^2 +
                        ({\theta } + \bar{\theta })
                       ]
        \; ,
\\
\label{kxk}
         k_{2} \!\cdot\! k_{4} = k_{1} \!\cdot\! k_{3}
               & = &
                        \frac{1}{3}
                      [
                        \mu^2 +
                        (\bar{\epsilon}{\theta }
                        + {\epsilon}\bar{\theta })
                       ]
        \; ,
\\
\nonumber
         k_{3} \!\cdot\! k_{4} = k_{1} \!\cdot\! k_{2}
               & = &
                        \frac{1}{3}
                      [
                        \mu^2 +
                        ({\epsilon}{\theta }
                        + \bar{\epsilon}\bar{\theta })
                       ]
        \; .
\end{eqnarray}
                The standard Mandelstam variables are related
                to
        $ \theta , \bar{\theta } $
                by
\begin{eqnarray}
\label{stue}
                s  =
                        \frac{2}{3}
                            [ 2 \mu^2
                        - (  \theta
                            + \bar{\theta }
                          ) ]
        \; ; \;
                u  =
                        \frac{2}{3}
                            [ 2 \mu^2
                        - ( \bar{\epsilon} \theta
                            + {\epsilon} \bar{\theta }
                          ) ]
        \; ; \;
                t  =
                        \frac{2}{3}
                            [ 2 \mu^2
                        - ( {\epsilon} \theta
                            + \bar{\epsilon} \bar{\theta }
                          ) ]
        \; ,
\end{eqnarray}
                the restriction
(\ref{hmass})
                being provided by the property of the cubic roots
        $$
           1 +  \epsilon +  \bar{\epsilon} = 0
                \;  .
        $$

                The above expressions show
                that in terms of the variable
        $ \theta $ ($\bar{\theta } $)
                the Mandelstam plane is viewed as
                a complex plane
                where all permutations of pions
                are realized by the
        $  2 \pi / 3 $
                rotations and the complex conjugation.
                When all Mandelstam variables are restricted
                to real values
                (only this very case will be
                considered below)
                the variables
(\ref{teate})
                satisfy the conjugation condition
        $ {\theta}^{*} = \bar{\theta} $
                and the quantities
\begin{equation}
\label{tRtI}
         {\theta}_{R} \equiv
               ( {\theta} +  \bar{\theta} ) / 2
          \; , \; \;
         {\theta}_{I} \equiv
               ( {\theta} - \bar{\theta} ) / (2i)
\end{equation}
                are real;
                it is useful to have their expressions
                in terms of the
        $  s $--channel
                CMF momentum
        $  {\bf k}_{s} $
                and the scattering angle
        $ \Theta_{s} $
                (in what follows the subscript will
                be omitted  if the
        $  s $--channel
                origin of a variable is unambiguous):
\begin{eqnarray}
\label{teRteI}
              \theta_{R}  =
                       - (
                           2 \mu^{2} + 3 {\bf k}^{2}
                         )
        \; ; \; \;
              \theta_{I}  =
                       (
                         \sqrt{3} {\bf k}^{2} \cos \Theta
                       )
        \; .
\end{eqnarray}

                There exist two symmetric invariants
                of permutations
                which might be built of
                the covariant variables
(\ref{teate}),
                namely
\begin{equation}
\label{VW}
          V  \equiv
                {\theta}   \bar{\theta}
          \; , \; \;
          W \equiv
                {\theta}^{3} + \bar{\theta}^{3}
          \; .
\end{equation}

                The invariance of variables
(\ref{VW})
                under the crossing transformations
                makes it attractive to use them
                for the purpose of the threshold expansion.
                Subtracting the values
        $  V_{0} = 4 \mu^4 $,
        $  W_{0} = - 16 \mu^6 $
                which the variables
(\ref{VW})
                obtain at all thresholds of cross reactions
                one gets the variables
\begin{equation}
\label{vw}
          v  \equiv V - V_{0}
          \; , \; \;
          w \equiv W - W_{0}
          \;
\end{equation}
                uniformly describing
                a deviation from all thresholds.
                For example,
                the variable
        $ v $
                has the same form
\begin{eqnarray}
\nonumber
             v & = &
                 3 [ {\bf k}^{2}_{s}
                     (
                     4 \mu^{2} + 3 {\bf k}^{2}_{s}
                     )
                      + {\bf k}^{4}_{s} \cos^{2} \Theta_{s}
                   ]
\\
\label{vk}
              & = &
                 3 [ {\bf k}^{2}_{u}
                     (
                     4 \mu^{2} + 3 {\bf k}^{2}_{u}
                     )
                      + {\bf k}^{4}_{u} \cos^{2} \Theta_{u}
                   ]
\\
\nonumber
              & = &
                 3 [ {\bf k}^{2}_{t}
                     (
                     4 \mu^{2} + 3 {\bf k}^{2}_{t}
                     )
                      + {\bf k}^{4}_{t} \cos^{2} \Theta_{t}
                   ]
        \;
\end{eqnarray}
                in CMF variables
        ($  {\bf k}_{s} $,
        $ \Theta_{s} $),
        ($  {\bf k}_{u} $,
        $ \Theta_{u} $),
        ($  {\bf k}_{t} $,
        $ \Theta_{t} $)
                of all cross channels.
                This is evident from the definition
(\ref{teate})
                and the following transformation properties
                of the crossing--covariant variables:
\begin{equation}
\label{trnc}
\begin{array}{cccl}
             ( s, u, t) & \to & ( t, s, u )
&
                 ,
\\
             ({\theta}, \bar{\theta} )
                & \to &
                         ( {\epsilon} \theta,
                             \bar{\epsilon} \bar{\theta } )
&
                  ;
\end{array}
\hspace{0.4cm}
\begin{array}{cccl}
             ( s, u, t) & \to & ( s, t, u )
&
                 ,
\\
             ({\theta}, \bar{\theta} )
                & \to &
                         ( \bar{\theta }, \theta )
&
                  .
\end{array}
\end{equation}
                The expression of
        $ w $
                in terms of
        $  {\bf k} $,
        $ \Theta $
                looks more involved,
\begin{equation}
\label{wk}
             w =
                -18 {\bf k}^{2}
                [
                  4 \mu^4 - 6 {\bf k}^{4}
                          + 3 {\bf k}^{2}
                               ( 1 - \cos^{2} \Theta )
                               ( 2 \mu^2 + 3 {\bf k}^{2} )
                ]
          \; ,
\end{equation}
                nevertheless, the invariance is obvious from
                the definition
(\ref{VW})
                and the transformation rules
(\ref{trnc}).

                The discussed invariance has the following
                important consequence:
                given a function
        $  F ( v, w ) $
                of the variables
        $ v $, $ w $
                only,
                it has the same partial wave (PW) expansion
                in all physical domains:
\begin{equation}
\label{PWF}
             F ( v, w ) =
             \sum_{L} ( 2 L + 1 )
                f_{L} ( {\bf k})
                        {\bf k}^{2L} P_{L} ( \cos \Theta)
          \; ,
\end{equation}
                where
        $  {\bf k} $,
        $ \Theta $
                can stand for variables of any channel
        $ s $,
        $ u $,
        $ t $,
                functions
        $ f_{L} $
                being the same.

                Now it is time to discuss
                the general structure
                of the threshold expansion of the isoscalar
                amplitudes.
                Because of relations
(\ref{crr})
                it is sufficient
                to consider only one amplitude,
                say,
        $ A $.
                Our purpose is to find structures
                which, like the nonanalytic terms of
                ChPT
                (Chiral logs),
                must be treated nonperturbatively.

                Let us assume for a moment
                that nonanalytic terms are absent
                and consider the
                ChPT
                expansion of
        $ A $
                in powers of pion momenta
        $ k_{1} $,
        $ k_{2} $,
        $ k_{3} $,
        $ k_{4} $.

                Being a Lorentz--invariant expression,
                amplitude is built of scalar products
                which by virtue of relations
(\ref{kxk})
                are functions of the independent
                crossing--covariant variables
(\ref{teate})
                only.
                Hence,
                ChPT
                expansion
                is an expansion in the variables
        $ {\theta} $, $ \bar{\theta} $.
                In respect to crossing transformations
                any such expansion splits into three
                parts:

                1) invariant terms built of
        $ {\theta}^{3} $,
        $ \bar{\theta}^{3} $
                and
        $ {\theta}  \bar{\theta} $;

                2) terms built of
        $ {\theta} $
                or
        $ \bar{\theta}^{2} $
                multiplied by invariant combination
                of point 1);
                these terms transform like
        $ {\theta} $;

                3) terms transforming like
        $ \bar{\theta} $ ---
                i.e. built of
        $ \bar{\theta} $
                or
        $ {\theta}^{2} $
                times invariant combinations.

                Because of properties
(\ref{crc})
                the number of distinct structures
                entering the particular amplitude
                is
                further reduced to three.
                By
(\ref{crc})
                the amplitude
        $ A $
                is a symmetric function
                of
        $ {\theta} $,
        $ \bar{\theta} $;
                hence, its expansion might be rewritten
                in terms of the sum and of the product of
                the arguments.
                The product is just
        $ V $
                variable,
                so, the expansion reads
\begin{equation}
\label{Arn}
             A ( {\theta}, \bar{\theta}  ) =
             \sum_{n}
                \alpha_{n} ( V )
                ( {\theta} + \bar{\theta} )^{n}
          \; ,
\end{equation}
                where coefficients
        $ \alpha_{n} $
                stand for expansions in
        $ V = V_{0} + v $.

                Now,
                the algebraic identity
\begin{equation}
\label{rid}
                ( {\theta} + \bar{\theta} )^{3}
        =
                3 ( {\theta} + \bar{\theta} )  V
                + W
\end{equation}
                helps to get rid of any powers of
        $ ( {\theta} + \bar{\theta} ) $
                greater than 2
                and to
                rewrite the amplitude
(\ref{Arn})
                in the form
\begin{equation}
\label{ArA}
             A ( {\theta}, \bar{\theta}  )
              =
                 A_{0} ( v , w )
              +
                ( {\theta} + \bar{\theta} )
                 A_{1} ( v , w )
              +
                ( {\theta} + \bar{\theta} )^{2}
                 A_{2} ( v , w )
          \;
\end{equation}
                or in an equivalent form,
                if the transformation properties
                discussed in the points 1), 2), 3)
                are of importance:
\begin{equation}
\label{Ara}
             A ( {\theta}, \bar{\theta}  )
              =
                 a_{0} ( v , w )
              +
                ( {\theta} + \bar{\theta} )
                 a_{1} ( v , w )
              +
                ( \bar{\theta}^{2} + {\theta}^{2} )
                 a_{2} ( v , w )
          \; .
\end{equation}
                Here,
                coefficients
        $ A_{0} $, $ A_{1} $, $ A_{2} $
        ($ a_{0} $, $ a_{1} $, $ a_{2} $)
                are crossing--invariant expansions.
                In the case of the amplitude
                which has no nonanalytic terms
                in the Mandelstam domain
(\ref{OMD})
                the coefficients might be reexpanded in
                the uniform threshold variables
        $ v $,
        $ w $.
                The
                expressions
        $ ( {\theta} + \bar{\theta} ) $,
        $ ( {\theta} + \bar{\theta} )^{2} $
                appears to be the natural
                structures
                for the decomposition of
                any analytic function.

                In the general case the
        $ \pi \pi $
                amplitude,
                being analytic in the domain
(\ref{OMD})
                investigated by Mandelstam,
                admits an expansion with the restricted
                convergence radius.
                The latter is determined by the location
                of amplitude singularities
                in the complex region of variables.
                The singularities include the branching
                points connected to the thresholds of
        $ s $,  $ t $,  $ u $
                channels of the considered reaction
                represented by
                the phase--space factor
        $ \sigma ({\bf k_{s}}^{2}) $
                entering the solution
(\ref{Imsol})
                (--- only these very singularities are of
                importance below the inelastic threshold).

                We intend to treat separately the real
        $ {\rm Re} A $
                and the imaginary part
        $ {\rm Im} A $
                of our amplitude;
                this ultimately provides the possibility
                to concentrate all free parameters
                in the real part
                (making the relation with observables,
                i.e. scattering lengths and slopes easy)
                and to use the perturbative unitarity
                for eliminating all parameters
                of the imaginary part.

                Let us define 3 quantities
\begin{eqnarray}
\label{rstu}
                r_{s}  =  \sqrt{  ( s/4 - \mu^{2}) s }
        \; ; \;\;
                r_{u}  =  \sqrt{  ( u/4 - \mu^{2}) u }
        \; ; \;\;
                r_{t}  =  \sqrt{  ( t/4 - \mu^{2}) t }
        \;
\end{eqnarray}
                to be positive at real values of
        $  s, t, u $
                corresponding to physical domains of all
                cross channels of the
        $  \pi \pi $
                reaction.
                We shall consider scattering lengths and
                slopes which are determined
                in the physical region only,
                so there will be no need in the exact global
                definition of phases of square roots.

                Since the powers of
        $ r_{s} $, $ r_{u} $, $ r_{t} $
                which are greater than 3 are being reduced
                to the set
        $ \{ r_{s} $, $ r_{u} $, $ r_{t} $;
        $ r_{s} r_{u} $, $ r_{u} r_{t} $, $ r_{t} r_{s} $;
        $ r_{s}  r_{u}  r_{t} \}$
                times  polynomials in
        $ {s} $, $ {u} $, $ {t} $
                and
                only odd powers of
        $ r_{s} $, $ r_{u} $, $ r_{t} $
                develop the singularity
        $  \sqrt{ {\bf k}^{2}} $
                which is characteristic to the imaginary part
                of the amplitude according to eq.
(\ref{ImTl})
                the general algebraic form of a function
                of these quantities is as follows:
\begin{equation}
\label{ImArsut}
             {\rm Im}~A  =
                r_{s}             F_{s} +
                r_{u}             F_{u} +
                r_{t}             F_{t} +
                r_{s} r_{u} r_{t} F_{0}
          \; .
\end{equation}
                Due to the
        ($ t \leftrightarrow u $)
                symmetry of the amplitude
        $ A $
                it is convenient to  rewrite
(\ref{ImArsut})
                in the form
\begin{equation}
\label{ImArsutsym}
             {\rm Im}~A  =
                r_{s}                         F_{s} +
                ( r_{u} + r_{t} )             F_{+} +
                i ( {\theta} - \bar{\theta} )
                  ( r_{u} - r_{t} )           F_{-} +
                r_{s} r_{u} r_{t}             F_{0}
          \; ,
\end{equation}
                where
        $ F_{N} $
        ($ N = s,+,-,0 $)
                are regular functions of
        $  { {\bf k}^{2}} $
                at the threshold
                (hence,
                they are regular in
        $ {\theta} - {\theta}_{0} $,
        $ \bar{\theta} - {\theta}_{0} $
                as well).
                Therefore,
                one might assume
                that the restricted convergence
                of an expansion of the amplitude
                is due to square--root singularities
        $ r_{s} $, $ r_{u} $, $ r_{t} $
                and the regular coefficient functions
        $ F_{N} $
                (or
        $  F_{s} $, $  F_{u} $,  $  F_{t} $,  $  F_{0} $)
                have the larger domain of convergence
                than
        $ {\rm Im}~A $
                itself.
                Then their expansions
                might be rewritten in the form
\begin{equation}
\label{FN}
             F_{N} ( {\theta}, \bar{\theta}  ) =
                 F_{N}^{0} ( v , w )
              +
                ( {\theta} + \bar{\theta} )
                F_{N}^{1} ( v , w )
              +
                ( \bar{\theta} + {\theta} )^{2}
                F_{N}^{2} ( v , w )
        \; \; \;
                ({N} = s, +, -, 0 )
          \;
\end{equation}
                where the crossing invariant functions
        $  F_{N}^{\nu} $,
        $  {\nu} = 0,1,2 $
                describe
        $  F_{N} $
                in all threshold regions simultaneously.

                One can see that the existence of additional
                variables
                (even with algebraically restricted powers
                to the lower ones only)
                significantly increases the number
                of degrees of freedom of a crossing
                covariant ansatz.

                Hopefully,
                because of unitarity conditions
                there appears no independent parameters
                at all
                in the imaginary part.
                Indeed,
                when the expansion
(\ref{FN})
                is being brought to
                the form
(\ref{ImTl})
                some combinations of coefficients
                must be set to zero
                while the rest become expressible
                in terms of the low energy parameters of
                the real part.

                Since we need the representation of the
        $ \pi \pi $
                amplitude
                which is suitable for the data analysis
                in the most simple terms
                we shall not discuss the properties
                of the expansion
(\ref{FN})
                any more.
                In what follows
                the suitable expression for the imaginary part
                will be obtained in a more straightforward
                way.

\newpage
      \section{ Near--Threshold Phenomenological Amplitude }

                We are able now to discuss the form of the
                threshold
        $  \pi \pi $
                amplitude.

                The strength of ChPT predictions relies upon
                the hypothesis that its
        ($  {\bf k}, \mu $)
                expansion is convergent
                not only in the central triangle
                of the Mandelstam plane
                but well above the thresholds
                of physical channels
                --- the threshold characteristics
(\ref{ReTl})
                are then being easily determined.
                Let us assume that
                at physical
                masses
                at least the real part of
                the phenomenological amplitude
                has the same property.
                Then the real part of the isoscalar amplitude
        $ A $
                is given by eq.
(\ref{ArA})
                of the previous section.
                The realization
(\ref{trnc})
                of the crossing transformations in relations
(\ref{crr})
                provides expressions
                for all 3 isoscalar amplitudes
\begin{eqnarray}
\nonumber
             {\rm Re} ~A
        & = &
                 A_{0} ( v , w )
              +
                ( {\theta} + \bar{\theta} )
                 A_{1} ( v , w )
              +
                ( {\theta} + \bar{\theta} )^{2}
                 A_{2} ( v , w )
                 \; , \\
\label{ReABC}
             {\rm Re} ~B
        & = &
                 A_{0} ( v , w )
              +
                ( \bar{\epsilon} {\theta}
                + {\epsilon} \bar{\theta} )
                 A_{1} ( v , w )
              +
                ( \bar{\epsilon} {\theta}
                + {\epsilon} \bar{\theta} )^{2}
                 A_{2} ( v , w )
                 \; , \\
\nonumber
             {\rm Re} ~C
        & = &
                 A_{0} ( v , w )
              +
                ( {\epsilon} {\theta}
                + \bar{\epsilon} \bar{\theta} )
                 A_{1} ( v , w )
              +
                ( {\epsilon} {\theta}
                + \bar{\epsilon} \bar{\theta} )^{2}
                 A_{2} ( v , w )
                 \; .
\end{eqnarray}
                Here,
                crossing invariant functions
        $ A_{\nu} ( v , w ) $,
        ($ {\nu} = 0, 1, 2  $)
                are expansions in the threshold variables
(\ref{vw})
((\ref{vk}), (\ref{wk}))
\begin{equation}
\label{Anu}
              A_{\nu} ( v, w ) =
                ( 32 \pi )
                        \sum_{m, n}
                        g_{m n}^{\nu} v^{m} w^{n}
                \;
                \;
                ( \nu = 0, 1, 2 )
                \; ,
\end{equation}
                determined by arrays of coefficients
        $ g_{m n}^{\nu} $.
                These coefficients which are free from
                isotopic and crossing constraints
                will be considered as the
                {\it independent phenomenological parameters}
                of the
        $  \pi \pi $
                amplitude.

                The expansion of functions
        $ A^{\nu} ( v, w ) $
                is assumed to be valid for
        $ v $
                and
        $ w $
                bounded by some values
        $ v_{1} $,
        $ w_{1} $
\begin{eqnarray}
\label{v1w1}
                0 \; \leq  v  \leq \; v_{1}
        \; ; \;\;
                w_{1} \; \leq  w  \leq \; 0
        \; .
\end{eqnarray}
                This domain is shown on the Fig. 1.,
                where the Mandelstam plane
                containing the
        $ s $,
        $ u $,
        $ t $
                physical regions is drawn.
                The curves
        $ v = 0 $,
        $ v = v_{1} $
                are circles;
                together with the cubic curves
        $ w = 0 $,
        $ w = w_{1} $
                (looking like hyperbolas)
                they form the lens--like domain
                containing a part of the physical region
                from the threshold
                up to some energy
        $ s_{1} $.
                In fact,
                because of the complete crossing invariance
                of the considered functions
                there are 3 such domains
                at thresholds of all physical channels
                in which these functions
                acquire the same values.


\begin{figure}[thb]
\centerline{\psfig{figure=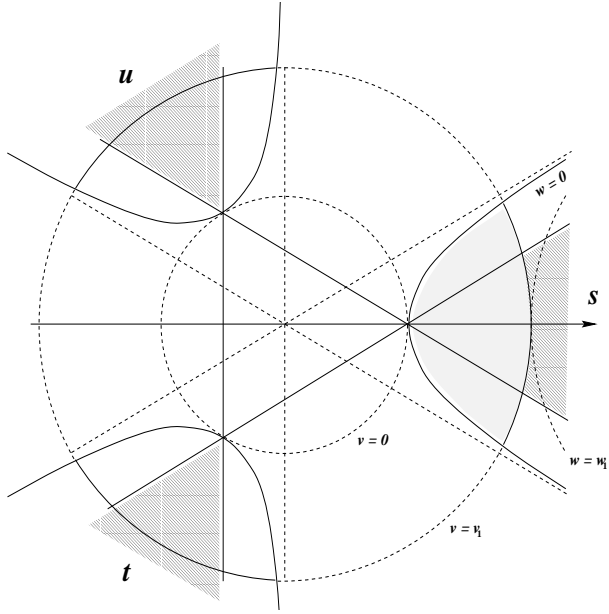,height=9cm,width=8cm,angle=-90}}
\caption{
\label{stufig}
Central region of Mandelstam plane.
                {\it
The filled area bounded by curves
        $ v = v_{1} $,
        $ w = 0 $
        is the expansion domain of the
        crossing invariant amplitudes
        $ A^{\nu} ( v, w ) $.
                }
         }
\end{figure}

                We already know
                (see eq.
(\ref{PWF}))
                that for such functions
                the Partial Wave Analysis (PWA)
                is identical in all physical regions.
                The coefficients
        $ A^{L,N}_{\nu} $
                of the PWA expansion
\begin{equation}
\label{PWAAnu}
              A_{\nu} ( v, w ) =
                       (32 \pi)
                        \sum_{L={\rm even}} ( 2 L + 1 )
                        A_{\nu}^{L} ({\bf k}^{2})
                        P_{L} ( \cos \Theta )
                \; ;
\end{equation}
\begin{equation}
\label{Anuk}
              A_{\nu}^{L} ({\bf k}^{2}) =
                        \sum_{N = 0}
                        {\bf k}^{2(N+L)} A_{\nu}^{L,N}
                \;
                \;
                ( \nu = 0, 1, 2 )
                \;
\end{equation}
                are easily calculated
                for every term of the expansion
(\ref{Anu})
                by the formula
\begin{equation}
\label{AnuLN}
              A_{\nu}^{L,N}   =
                        \frac{1}{ 64 \pi }
                        \frac{1}{ (N+L) ! }
                        \left (
                        \frac{d}{ d ({\bf k}^{2}) }
                        \right )^{N+L}
                \int_{-1}^{1} d \cos \Theta
                        ~P_{L} ( \cos \Theta )
                    A_{\nu} ( v, w )
                        ~\vert_{{\bf k}=0}
                \; ,
\end{equation}
                where expressions for variables
        $ v $,
        $ w $
                in terms of
        $ {\bf k}^{2} $,
        $ \cos \Theta $
                are given by relations
(\ref{vk}),
(\ref{wk}).

                For 6 lowest terms of the expansion
(\ref{Anu})
\begin{equation}
\label{sixAnunm}
              A^{\nu} ( v, w )
                =
                       (32 \pi)
                (
                     g^{\nu}_{0 0}
                +
                     v g^{\nu}_{1 0}
                +
                     w g^{\nu}_{0 1}
                +
                     v^{2} g^{\nu}_{2 0}
                +
                     v w g^{\nu}_{1 1}
                +
                     w^{2} g^{\nu}_{0 2}
                +   \dots
                )
                \;
\end{equation}
                the nonzero PWA parameters are
\begin{eqnarray}
\label{Anu00}
A_{\nu}^{0,0} & = & g^{\nu}_{0 0}
\; ;
\\
\label{Anu01}
A_{\nu}^{0,1} & = & 12 (g^{\nu}_{1 0} - 6 g^{\nu}_{0 1})
\; ;
\\
\label{Anu02}
A_{\nu}^{0,2} & = & 2 (72 g^{\nu}_{2 0} - 432 g^{\nu}_{1 1}
                + 5 g^{\nu}_{1 0} + 2592 g^{\nu}_{0 2}
                - 48 g^{\nu}_{0 1})
\; ;
\\
\label{Anu03}
A_{\nu}^{0,3} & = & 12 (20 g^{\nu}_{2 0} - 156 g^{\nu}_{1 1}
                + 1152 g^{\nu}_{0 2} - 3 g^{\nu}_{0 1})
\; ;
\\
\label{Anu04}
A_{\nu}^{0,4} & = & 72 (7 g^{\nu}_{2 0} - 96 g^{\nu}_{1 1}
                + 1008 g^{\nu}_{0 2})/5
\; ;
\\
\label{Anu05}
A_{\nu}^{0,5} & = & 1728 ( - g^{\nu}_{1 1} + 21 g^{\nu}_{0 2})
                /5
\; ;
\\
\label{Anu06}
A_{\nu}^{0,6} & = & 7776 g^{\nu}_{0 2}/5
\; ;
\\
\label{Anu20}
A_{\nu}^{2,0} & = & 2 (g^{\nu}_{1 0} + 12 g^{\nu}_{0 1})
                /5
\; ;
\\
\label{Anu21}
A_{\nu}^{2,1} & = & 12 (4 g^{\nu}_{2 0} + 12 g^{\nu}_{1 1}
                - 288 g^{\nu}_{0 2} + 3 g^{\nu}_{0 1})
                /5
\; ;
\\
\label{Anu22}
A_{\nu}^{2,2} & = & 288 (g^{\nu}_{2 0} + 12 g^{\nu}_{1 1}
                - 234 g^{\nu}_{0 2})/35
\; ;
\\
\label{Anu23}
A_{\nu}^{2,3} & = & 432 (5 g^{\nu}_{1 1} - 132 g^{\nu}_{0 2})
                /35
\; ;
\\
\label{Anu24}
A_{\nu}^{2,4} & = &  - 15552 g^{\nu}_{0 2}/35
\; ;
\\
\label{Anu40}
A_{\nu}^{4,0} & = & 8 (g^{\nu}_{2 0} + 12 g^{\nu}_{1 1}
                + 144 g^{\nu}_{0 2})/35
\; ;
\\
\label{Anu41}
A_{\nu}^{4,1} & = & 144 (g^{\nu}_{1 1} + 24 g^{\nu}_{0 2})
                /35
\; ;
\\
\label{Anu42}
A_{\nu}^{4,2} & = & 2592 g^{\nu}_{0 2}/35
\;  .
\end{eqnarray}

                It is reasonable to consider
                the coefficients
        $  A_{\nu}^{L,N} $
                as
                the {\it subsidiary} phenomenological
                parameters.
                These parameters are directly related to
                the ordinary threshold parameters
                of the
        $  \pi \pi $
                amplitude
                entering eq.
(\ref{ReTl}).
                For partial waves of isospin amplitudes
\begin{equation}
\label{TI0}
             T^{I=0} = 3 A + B + C =
                        5 A_{0} + 4 \theta_{R} A_{1}
                        + 2 [ (2 \theta_{R})^{2} + 3 V ] A_{2}
                \; ;
\end{equation}
\begin{equation}
\label{TI1}
             T^{I=1} =   - B + C =
                -2 \sqrt{3} \theta_{I}
                        \{ A_{1} - 2 \theta_{R} A_{2} \}
                \; ;
\end{equation}
\begin{equation}
\label{TI2}
             T^{I=2} =    B + C =
                        2 A_{0} - 2 \theta_{R} A_{1}
                        - [ (2 \theta_{R})^{2} - 6 V ] A_{2}
                \;
\end{equation}
                one can find the following relations:
\begin{eqnarray}
\nonumber
             T_{L}^{I=0}
    & = &
           5  A_{0}^{L} + 4 \theta_{R} A_{1}^{L}
           + 2
                    \left [
                       5 \theta_{R}^{2}
                        +
                        9 {\bf k}^{4}
                        \frac   {2 L^{2} + 2L - 1}
                               {(2L-1)(2L+3)}
                \right ]
              A_{2}^{L}
\\
\label{TI0l}
    &  &
                        +
                        18 {\bf k}^{4}
                \left [
                        \frac   {L-1}
                               {2L-1}
                        \frac   {L}
                               {2L+1}
                        A_{2}^{L-2}
                        +
                        \frac   {L+1}
                               {2L+1}
                        \frac   {L+2}
                               {2L+3}
                        A_{2}^{L+2}
                \right ]
                \; ;
\end{eqnarray}
\begin{equation}
\label{TI1l}
             T_{L}^{I=1} =
                       \sqrt{3} {\bf k}^{2}
                \left [
                        \frac   {L}
                               {2L+1}
                         (
                        A_{1}^{L-1} - 2 \theta_{R} A_{2}^{L-1}
                         )
                        +
                        \frac   {L+1}
                               {2L+1}
                         (
                        A_{1}^{L+1} - 2 \theta_{R} A_{2}^{L+1}
                         )
                \right ]
                \; ;
\end{equation}
\begin{eqnarray}
\nonumber
             T_{L}^{I=2}
    & = &
           2  A_{0}^{L} - 2 \theta_{R} A_{1}^{L}
           + 2
                    \left [
                       \theta_{R}^{2}
                        +
                        9 {\bf k}^{4}
                        \frac   {2 L^{2} + 2L - 1}
                               {(2L-1)(2L+3)}
                \right ]
              A_{2}^{L}
\\
\label{TI2l}
    &  &
                        +
                        18 {\bf k}^{4}
                \left [
                        \frac   {L-1}
                               {2L-1}
                        \frac   {L}
                               {2L+1}
                        A_{2}^{L-2}
                        +
                        \frac   {L+1}
                               {2L+1}
                        \frac   {L+2}
                               {2L+3}
                        A_{2}^{L+2}
                \right ]
                \; .
\end{eqnarray}
                In the above expressions
                whenever the partial--wave index
        $ L $
                of
        $ A_{\nu}^{L} $
                becomes less then zero
                the value of
        $ A_{\nu}^{L} $
                must be set to zero.
                Here, for the purpose of brevity the expression
(\ref{teRteI})
                for
        $ \theta_{R} = \theta_{R} ( {\bf k}^{2} ) $
                had not been expanded.

                The low energy parameters
(\ref{ReTl})
                are then directly expressed
                in terms of quantities
        $ \{ A_{\nu}^{L,N} \} $
                and, finally,
                in terms of the independent phenomenological
                parameters
        $ g_{m n}^{\nu} $
                of eq.
(\ref{Anu}).
                Leaving only the linear terms of the expansion
(\ref{sixAnunm})
                we get:
\begin{eqnarray}
\label{aI000}
{\bf [0]}
\hspace{0.5cm}
a^{I=0}_{0,0} & = & 5 g^0_{0 0} - 8 g^1_{0 0} {\mu}^2
        + 56 g^2_{0 0} {\mu}^4
\; ;
\\
{\bf [1]}
\hspace{0.5cm}
a^{I=0}_{0,1} & = & 12 (5 g^0_{1 0} {\mu}^2
        - 30 g^0_{0 1} {\mu}^4 - 8 g^1_{1 0} {\mu}^4
        + 48 g^1_{0 1} {\mu}^6 - g^1_{0 0}
        + 56 g^2_{1 0} {\mu}^6
\nonumber
\\
 & &
        - 336 g^2_{0 1} {\mu}^8 + 14 g^2_{0 0} {\mu}^2)
\; ;
\\
{\bf (2)}
\hspace{0.5cm}
a^{I=0}_{0,2} & = & 2 (25 g^0_{1 0} - 240 g^0_{0 1} {\mu}^2
        - 112 g^1_{1 0} {\mu}^2 + 816 g^1_{0 1} {\mu}^4
        + 1288 g^2_{1 0} {\mu}^4
\nonumber
\\
 & &
        - 8736 g^2_{0 1} {\mu}^6 + 66 g^2_{0 0})
\; ;
\\
{\bf (3)}
\hspace{0.5cm}
a^{I=0}_{0,3} & = & 12 ( - 15 g^0_{0 1} - 10 g^1_{1 0}
        + 120 g^1_{0 1} {\mu}^2 + 272 g^2_{1 0} {\mu}^2
        - 2304 g^2_{0 1} {\mu}^4)
\; ;
\\
{\bf (4)}
\hspace{0.5cm}
a^{I=0}_{0,4} & = & 144 (15 g^1_{0 1} + 46 g^2_{1 0}
        - 648 g^2_{0 1} {\mu}^2) / 5
\; ;
\\
{\bf (5)}
\hspace{0.5cm}
a^{I=0}_{0,5} & = &  - 23328 g^2_{0 1} / 5
\; ;
\\
{\bf [2]}
\hspace{0.5cm}
a^{I=0}_{2,0} & = & 2 (5 g^0_{1 0} + 60 g^0_{0 1} {\mu}^2
        - 8 g^1_{1 0} {\mu}^2 - 96 g^1_{0 1} {\mu}^4
        + 56 g^2_{1 0} {\mu}^4
\nonumber
\\
 & &
        + 672 g^2_{0 1} {\mu}^6 + 6 g^2_{0 0}) / 5
\; ;
\\
{\bf (3)}
\hspace{0.5cm}
a^{I=0}_{2,1} & = & 12 (15 g^0_{0 1} - 2 g^1_{1 0}
        - 48 g^1_{0 1} {\mu}^2 + 40 g^2_{1 0} {\mu}^2
        + 432 g^2_{0 1} {\mu}^4) / 5
\; ;
\\
{\bf (4)}
\hspace{0.5cm}
a^{I=0}_{2,2} & = & 144 ( - 21 g^1_{0 1} + 19 g^2_{1 0}
        + 396 g^2_{0 1} {\mu}^2) / 35
\; ;
\\
{\bf (5)}
\hspace{0.5cm}
a^{I=0}_{2,3} & = & 31104 g^2_{0 1} / 35
\; ;
\\
{\bf (4)}
\hspace{0.5cm}
a^{I=0}_{4,0} & = & 48 (g^2_{1 0} + 12 g^2_{0 1} {\mu}^2) / 35
\; ;
\\
{\bf (5)}
\hspace{0.5cm}
a^{I=0}_{4,1} & = & 864 g^2_{0 1} / 35
\; ;
\\
{\bf [1]}
\hspace{0.5cm}
a^{I=1}_{1,0} & = & 2 ( - g^1_{0 0} + 4 g^2_{0 0} {\mu}^2)
\; ;
\\
{\bf [2]}
\hspace{0.5cm}
a^{I=1}_{1,1} & = & 12 ( - 2 g^1_{1 0} {\mu}^2
        + 12 g^1_{0 1} {\mu}^4 + 8 g^2_{1 0} {\mu}^4
        - 48 g^2_{0 1} {\mu}^6 + g^2_{0 0})
\; ;
\\
{\bf (3)}
\hspace{0.5cm}
a^{I=1}_{1,2} & = & 36 ( - 3 g^1_{1 0}
        + 24 g^1_{0 1} {\mu}^2 + 32 g^2_{1 0} {\mu}^2
        - 216 g^2_{0 1} {\mu}^4) / 5
\; ;
\\
{\bf (4)}
\hspace{0.5cm}
a^{I=1}_{1,3} & = & 216 (g^1_{0 1} + 3 g^2_{1 0}
        - 28 g^2_{0 1} {\mu}^2) / 5
\; ;
\\
{\bf (5)}
\hspace{0.5cm}
a^{I=1}_{1,4} & = &  - 1296 g^2_{0 1} / 5
\; ;
\\
{\bf [3]}
\hspace{0.5cm}
a^{I=1}_{3,0} & = & 36 ( - g^1_{1 0} - 12 g^1_{0 1} {\mu}^2
        + 4 g^2_{1 0} {\mu}^2 + 48 g^2_{0 1} {\mu}^4) / 35
\; ;
\\
{\bf (4)}
\hspace{0.5cm}
a^{I=1}_{3,1} & = & 216 ( - 3 g^1_{0 1} + g^2_{1 0}
        + 24 g^2_{0 1} {\mu}^2) / 35
\; ;
\\
{\bf (5)}
\hspace{0.5cm}
a^{I=1}_{3,2} & = & 3888 g^2_{0 1} / 35
\; ;
\\
{\bf [0]}
\hspace{0.5cm}
a^{I=2}_{0,0} & = & 2 (g^0_{0 0} + 2 g^1_{0 0} {\mu}^2
        + 4 g^2_{0 0} {\mu}^4)
\; ;
\\
{\bf [1]}
\hspace{0.5cm}
a^{I=2}_{0,1} & = & 6 (4 g^0_{1 0} {\mu}^2
        - 24 g^0_{0 1} {\mu}^4 + 8 g^1_{1 0} {\mu}^4
        - 48 g^1_{0 1} {\mu}^6 + g^1_{0 0}
        + 16 g^2_{1 0} {\mu}^6
\nonumber
\\
 & &
        - 96 g^2_{0 1} {\mu}^8 + 4 g^2_{0 0} {\mu}^2)
\; ;
\\
{\bf (2)}
\hspace{0.5cm}
a^{I=2}_{0,2} & = & 4 (5 g^0_{1 0} - 48 g^0_{0 1} {\mu}^2
        + 28 g^1_{1 0} {\mu}^2 - 204 g^1_{0 1} {\mu}^4
        + 92 g^2_{1 0} {\mu}^4
\nonumber
\\
 & &
        - 624 g^2_{0 1} {\mu}^6 + 6 g^2_{0 0})
\; ;
\\
{\bf (3)}
\hspace{0.5cm}
a^{I=2}_{0,3} & = & 12 ( - 6 g^0_{0 1} + 5 g^1_{1 0}
        - 60 g^1_{0 1} {\mu}^2 + 44 g^2_{1 0} {\mu}^2
        - 360 g^2_{0 1} {\mu}^4)
\; ;
\\
{\bf (4)}
\hspace{0.5cm}
a^{I=2}_{0,4} & = & 72 ( - 15 g^1_{0 1} + 17 g^2_{1 0}
        - 216 g^2_{0 1} {\mu}^2) / 5
\; ;
\\
{\bf (5)}
\hspace{0.5cm}
a^{I=2}_{0,5} & = &  - 3888 g^2_{0 1} / 5
\; ;
\\
{\bf [2]}
\hspace{0.5cm}
a^{I=2}_{2,0} & = & 4 (g^0_{1 0} + 12 g^0_{0 1} {\mu}^2
        + 2 g^1_{1 0} {\mu}^2 + 24 g^1_{0 1} {\mu}^4
        + 4 g^2_{1 0} {\mu}^4
\nonumber
\\
 & &
        + 48 g^2_{0 1} {\mu}^6 + 3 g^2_{0 0}) / 5
\; ;
\\
{\bf (3)}
\hspace{0.5cm}
a^{I=2}_{2,1} & = & 12 (6 g^0_{0 1} + g^1_{1 0}
        + 24 g^1_{0 1} {\mu}^2 + 16 g^2_{1 0} {\mu}^2) / 5
\; ;
\\
{\bf (4)}
\hspace{0.5cm}
a^{I=2}_{2,2} & = & 72 (21 g^1_{0 1} + 17 g^2_{1 0}
        + 36 g^2_{0 1} {\mu}^2) / 35
\; ;
\\
{\bf (5)}
\hspace{0.5cm}
a^{I=2}_{2,3} & = & 3888 g^2_{0 1} / 35
\; ;
\\
{\bf (4)}
\hspace{0.5cm}
a^{I=2}_{4,0} & = & 48 (g^2_{1 0} + 12 g^2_{0 1} {\mu}^2) / 35
\; ;
\\
\label{aI241}
{\bf (5)}
\hspace{0.5cm}
a^{I=2}_{4,1} & = & 864 g^2_{0 1} / 35
\;  .
\end{eqnarray}

                These relations establish the connection
                between the standard low energy parameters
                which are nonzero in the linear in
        $ v $,
        $ w $
                approximation for all invariant functions
        $ A_{\nu} $
                and the
                real part of the phenomenological amplitude
                defined by eqs.
(\ref{ReABC}),
(\ref{Anu}).
                Here,
                the bold--face numbers in the LHS
                show the order in
        $ {\bf k}^{2} $;
                the numbers in the square brackets
                refer to the quantities
                which do not undergo corrections
                when three rest terms of the expansion
(\ref{sixAnunm})
                (as well as all higher order terms)
                are added.

                Before introducing
                the form for the imaginary part
                let us briefly discuss
                the specific features
                of the considered ansatz.

                First,
                one can find that 3 lowest
                scattering lengths
        $ a^{I=0}_{0,0} $,
        $ a^{I=1}_{1,0} $,
        $ a^{I=2}_{0,0} $
                appear to be unconstrained
                already in the zero approximation.
                (For simplicity we do not count the powers
                coming from {\it structural} combinations
        $ ( {\theta} + \bar{\theta} ) $,
        $ ( {\theta} + \bar{\theta} )^{2} $,
        etc. in eqs.
(\ref{ReABC}).)
                The linear approximation
                with 9 phenomenological parameters
                determine the quantities
                which by an accident are also 9 in total
                (those marked with order value
                given in the square brackets).
                The rest scattering lengths and slopes
                can not be free:
                their total number is growing twice faster
                than the number of free phenomenological
                parameters.
                The analysis of relations
(\ref{TI0l}),
(\ref{TI1l}),
(\ref{TI2l})
                makes it evident
                that the standard low energy characteristics
        $ a^{I}_{L,N} $
                of the expansion
(\ref{ReTl})
                are in the one--to--one correspondence with
                the subsidiary parameters
        $ A^{L,N}_{\nu} $
                of the crossing invariant amplitudes
        $ A_{\nu} $.
                Therefore,
                it is the PWA expansion
(\ref{PWAAnu})
                where parameters
        $ A^{L,N}_{\nu} $
                can not be arbitrary:
                an inspection of the relations
(\ref{Anu00})--(\ref{Anu42})
                might illustrate this.
                In  the general case,
                bringing an expansion of the kind
(\ref{PWAAnu})
                back to the form
(\ref{ArA})
                of the previous section
                one must
                obtain the vanishing coefficients at
        $ ( {\theta} + \bar{\theta} ) $
                and
        $ ( {\theta} + \bar{\theta} )^{2} $
                structures
                because of the crossing invariance of
                the considered function.
                Being expressed via subsidiary
                parameters
        $ A^{L,N}_{\nu} $
                and, finally, via scattering lengths
                and slopes
                these two vanishing coefficients
                provide two infinite sets of conditions
                for the low energy characteristics of the
        $  \pi \pi $
                amplitude.

                In the present work we do not plan to
                discuss
                in more details
                the arising conditions.
                We only would like to state
                that the conditions definitely differ
                from Roy equations
\cite{Roy71}
                since only the threshold characteristics
                are being involved.

                Second,
                one can derive more strong conclusions
                when invoking the properties of the
        $  \pi \pi $
                scattering
                implied by ChPT.
                This time the structure
        $ ( {\theta} + \bar{\theta} ) $
        ($ ( {\theta} + \bar{\theta} )^{2} $)
                must be considered as the
        $ O ( k^{2} ) $
        ($ O ( k^{4} ) $)
                quantity.
                What is more important
                it is the increased ChPT weight
                of basic variables:
        $ v  \sim O ( k^{4} ) $,
        $ w  \sim O ( k^{6} ) $.
                This results in the
                different counting scheme
                which, for example, at
        $ O ( k^{6} ) $
                leaves in effect only
                the following parameters of the
                considered approximation:
\begin{equation}
\label{ChPTorder}
\begin{array}{ccccl}
       O ( k^{0} ) \; : \hspace{0.3cm}  &
                         g^0_{0 0}  &
                                         &
&  ;
\\
       O ( k^{2} ) \; : \hspace{0.3cm}  &
                      &  g^1_{0 0}  &
&  ;
\\
       O ( k^{4} ) \; : \hspace{0.3cm}  &
                         g^0_{1 0}  &
                      &  g^2_{0 0}
&  ;
\\
       O ( k^{6} ) \; : \hspace{0.3cm}  &
                         g^0_{0 1}  &
                         g^1_{1 0}  &
&  .
\end{array}
\end{equation}
                Since no dynamics was used in the above
                pure kinematical considerations
                the actual scheme might be even more
                constrained.
                For example,
                to make the amplitude vanish in the
                chiral limit the value of
        $ g^0_{0 0} $
                must be of the order
        $ O ( \mu^{2} ) $.
                (In the leading order of ChPT one has
        $ g^0_{0 0} = \mu^2 / (3 F^{2}_{\pi})$,
        $ g^1_{0 0} = -2 / (3 F^{2}_{\pi}) $.)

                However, to provide the model--independent
                test of ChPT predictions one should not
                rely upon the above scheme.
                This does not mean that one inavoidably
                has to proceed with the third order
                expansion
(\ref{Anu})
                for testing the
                two--loop results.
                Since the latter are given as corrections
                to the quantities of the lower order
                the linear approximation in the expansion
(\ref{Anu})
                is sufficient both for testing
                the two--loop calculations
                and for confronting the standard
                and the generalized
                ChPT predictions
\cite{BijenesCEGS96,KnechtMSF95}.

                At last, one must note the rather large
                degree of slope parameters and higher
                scattering lengths
                which are necessary to be present in the
                amplitude to ensure the balance of the
                crossing properties
                --- this is illustrated by relations
(\ref{Anu00})--(\ref{Anu42}),
(\ref{aI000})--(\ref{aI241}).
                The neglect of such a parameter,
                if at all possible without prescribing
                definite value for a lower one,
                causes the appearance of higher terms
                in the expansion
(\ref{Anu})
                for the compensation.

                To finish the amplitude development we must
                consider the imaginary part.
                The simple model
(\ref{ImArsut}),
(\ref{ImArsutsym})
                of the previous section
                provides an illustration of the fact
                that there are no constraints
                on expansion coefficients
                of the imaginary part
                originating from the pure crossing symmetry.
                On the other hand the imaginary part
                is known to be completely determined by
                the unitarity conditions provided the real
                part of the amplitude is given (cf. eq.
(\ref{ImTlRe})).
                Since the applicability of the perturbative
                solution of unitarity
(\ref{Imsol})
                is limited by the condition
        $ | {\rm Re}~T^{I}_{L} | \approx | {\rm Im}~T^{I}_{L} | $
                and the only known phase shift satisfying
        $ \tan~\delta^{I}_{L} \approx \pm \pi / 4 $
                is the
        $ \delta^{I=0}_{0} $ one
                (at
        $ s \approx (4 \mu)^{2} $)
                we are safe to derive coefficients
                of the expansion
(\ref{ImTl})
                from eq.
(\ref{ImTlRe})
                below the inelastic threshold
                in the domain
(\ref{OMD}).
                The straightforward calculation results
                in the following expressions:
\begin{eqnarray}
\label{beta000}
\beta^{I=0,2}_{0,0} & = & (a^I_{0,0})^2
\; ;
\\
\beta^{I=0,2}_{0,1} & = & a^I_{0,0} (2 a^I_{0,1} {\mu}^2
        + (a^I_{0,0})^3)/{\mu}^2
\; ;
\\
\beta^{I=0,2}_{0,2} & = & (2 a^I_{0,2} a^I_{0,0} {\mu}^4
        + (a^I_{0,1})^2 {\mu}^4
        + 4 a^I_{0,1} (a^I_{0,0})^3 {\mu}^2
        + 2 (a^I_{0,0})^6 - (a^I_{0,0})^4)/{\mu}^4
\; ;
\\
\beta^{I=0,2}_{0,3} & = & (2 a^I_{0,3} a^I_{0,0} {\mu}^6
        + 2 a^I_{0,2} a^I_{0,1} {\mu}^6
        + 4 a^I_{0,2} (a^I_{0,0})^3 {\mu}^4
        + 6 (a^I_{0,1})^2 (a^I_{0,0})^2 {\mu}^4
\nonumber
\\
 & &
        + 12 a^I_{0,1} (a^I_{0,0})^5 {\mu}^2
        - 4 a^I_{0,1} (a^I_{0,0})^3 {\mu}^2
        + 5 (a^I_{0,0})^8 - 4 (a^I_{0,0})^6
        + (a^I_{0,0})^4)/{\mu}^6
\; ;
\\
\beta^{I=1}_{1,0} & = & (a^I_{1,0})^2
\; ;
\\
\label{beta111}
\beta^{I=1}_{1,1} & = & 2 a^I_{1,1} a^I_{1,0}
\;  .
\end{eqnarray}
                Here,
                we limited ourselves by few terms
                because higher waves are heavily
                suppressed at the threshold
                (cf. eq.
(\ref{ImTl})).
                As a result all parameters
        $ \beta^I_L  $
                of the imaginary part via relations
(\ref{aI000}),
(\ref{aI241})
                are determined in terms of
                the phenomenological parameters
        $ g_{m n}^{\nu} $
                of the expansion
(\ref{Anu});
                the expressions do not contain
                kinematics and allow fast fitting
                procedures when quantities
        $ \beta^I_L $
                are used as formal (dependent)
                parameters.

                This construction determines the
                imaginary parts of amplitudes
        $ T^{I} $
                with the
                fixed isospin of the decomposition
(\ref{MProj})
                in the
        $ s $--channel
                physical region.
                The isoscalar amplitudes in the same
                domain
                as well as any amplitude of a specific
                process are then simply determined
                by inverting the relations
(\ref{FxdI}).
                This completes the construction of the
                phenomenological
        $  \pi \pi $
                amplitude suitable for the analysis of
                experimental data in the low energy
                region.

      \section{
                Conclusions
              }

                We derived the general
                model--independent form
                of the real part of the
        $  \pi \pi $
                amplitude
                near the threshold
                basing on the nice invariance properties
                of the variables
        $  v, w $
                under the crossing transformations
                and the independence
                of the basic crossing--covariant variables
        $  {\theta} $, $ \bar{\theta} $.
                The expansion which we elaborated
                contains an equivalent of the threshold
                characteristics
                (scattering lengths and slopes)
                in an explicitly crossing covariant terms
                and it is valid for
                the threshold regions of all
                3 cross channels
        ($  s, t, u $)
                simultaneously.

                The advantage of the approach
                might be clearly displayed by a comparison
                with the analysis by Roskies
\cite{Roskies69}
                which is based on the ordinary
                Mandelstam variables
        $s, t, u$
                and is valid for the central point
        $s = t = u = 4 \mu^2 / 3 $.
                The sophisticated solution of the crossing
                restrictions in Mandelstam variables
                is given there in terms of the
                orthogonal polynomials
                over the central triangle
                of the Mandelstam plane
\cite{BalachandranN68}.
                The origin
                of the difference stems from the fact
                that in terms of the ordinary variables
                it is possible to construct only one
                crossing--invariant amplitude
                (namely,
        $ ( A + B + C ) / 3 = A^{0} + 2 V A^{2} $).

                Unfortunately,
                because of large errors of the existing data
                on the low energy
        $  \pi \pi $
                scattering
                now it is not possible
                to fix with the satisfactory precision
                the model--independent phenomenological
                parameters of our amplitude
(\ref{ReABC}),
(\ref{Anu}).
                For example,
                only four parameters are sufficient
                to fit (with
        $ \chi^{2} / N_{\rm DF} = 0.10 $)
                the data on scattering lengths and slopes
                of the compilation
\cite{Dumbrais83}
                providing (in the pion--mass units)
\begin{eqnarray}
\nonumber
          g^{0}_{00}  = & \; \; 0.024 \pm 0.005 \; \;
        \; ; \;\;
          g^{0}_{10} = & 0.00124 \pm 0.00025
        \; ;
\\
\label{g0g3}
          g^{1}_{00} = & -0.0178 \pm 0.0008
        \; ; \;\;
          g^{2}_{00} = & -0.00031 \pm  0.00015
        \; .
\end{eqnarray}

                Thus it is instructive to compare directly
                the theoretical amplitudes themselves,
                namely,
                the ChPT one--loop amplitude
        $  A_{\rm GL} $
                of the papers
\cite{GL8485}
                with that given by eqs.
(\ref{ReABC}),
(\ref{Anu})
                in the
        $  O ({\bf k}^{4}) $
                order.
                Minimizing the expression
\begin{equation}
\label{CompGL}
             \Delta =
              \frac
              {
                      \int_{4}^{8} ds \sigma (s)
                      \int_{4-s}^{0} dt
                \left |
                        A - A_{\rm GL}
                \right |^{2}
              }
              {
                      \int_{4}^{8} ds \sigma (s)
                      \int_{4-s}^{0} dt
                \left |
                         A_{\rm GL}
                \right |^{2}
              }
                \; ,
\end{equation}
                with the four--parameter amplitude
                one gets
        $  \Delta \simeq 0.0090  $.
                This is quite below the level of the accuracy
                as of the existing data
                as well as of the data which are being awaited
                in the near future
\cite{Pocanic94}.
                For simplicity only the real
                phenomenological amplitude was used.
                Since more realistic amplitude as well
                as the one of the higher ChPT order
                will have the imaginary part which
                might be only few times larger,
                the value of
        $  \Delta  $
                presents, in fact, an estimate to what
                extent the parameters
(\ref{beta000})--(\ref{beta111})
                of the imaginary part are effective
                in the Mandelstam domain
(\ref{OMD}).

                A brief summary of the most important
                properties of the discussed
                phenomenological amplitude includes:

                1. The Lorentz invariance of expressions
                determining the amplitude in terms of the
                independent parameters
                (in contrast to the explicit
                Reference--Frame dependence of the standard
                set in the definition
(\ref{ReTl})
                which complicates the application to processes
                like
        $  \pi N \to \pi \pi N $,
        $  \gamma N \to \pi \pi N $
                where the CMF
                of the dipion system does not coincide
                with the CMF of the reaction and, moreover,
                is ``moving'' from point to point
                of the phase space).

                2. The explicit crossing covariance
                and the universal description
                of the threshold regions
                of all 3 cross channels.

                3. Easy (perturbative) unitarity
                corrections
                which at the threshold
                always come in the higher order.

                Therefore,
                the presented phenomenological
        $  \pi \pi  $
                amplitude is suitable for the analysis of the
        $  K \to \pi \pi e {\nu} $
                decay  and the
        $  \pi N \to \pi \pi N $
                reaction from the threshold
        $  P_{\rm l.s. } = 280 $Mev/c
                up to
        $  P_{\rm l.s. } \simeq 500 $Mev/c
                since the energy region
        $  s_{\pi\pi} \sim 8 \mu^2  $
                is statistically insignificant or unreachable
                in the above conditions.
                The results then might find the application
                for the determination of the axial anomaly
                (when processing the final--state interaction
                corrections
                to the
        $  \gamma \to 3 \pi $
                vertex
                of the amplitude of
        $  \gamma N \to \pi \pi N $
                reactions)
                and in other similar cases.

                This research was supported in part
                by the RFBR grant N 95-02-05574a.

\end{document}